\title {Thoughts on Noise and Quantum
Computation\footnote{Research supported in part by an NSF grant,
by an ISF Bikura grant and by a BSF grant.
Part of this work was carried out when the author visited the
Mittag-Lefler Institute in Djursholm, Sweden.
I am  very thankful to Dorit Aharonov, Robert Alicki, 
Michael Ben-Or, Greg Kuperberg,
and Boris Tsirelson for helpful discussions, and to 
Daniel Gottesman, 
Pierfrancesco La Mura, Nati Linial, Simon Litsyn, Yuval
Peres, Itamar Pitowsky, Nick Read, Muli Safra, Oded Schramm, 
Anatoly Vershik, and
Avi Wigderson for helpful remarks.}
} 
\author {  Gil Kalai \\
Hebrew University of Jerusalem and Yale University}
\documentstyle[12pt,amsfonts]{article}
\newtheorem{conj}{Conjecture}[section]
\newtheorem{prob}[conj]{Problem}

\newcommand{\beq}[1]{\begin{equation}\label{#1}}
\newcommand{\enq}[0]{\end{equation}}

\renewcommand{\baselinestretch}{1.23}

\begin{document}

\maketitle

\begin {abstract}

We will try to 
%understand, 
explore, primarily from  the 
complexity-theoretic point of view, limitations of error-correction and
fault-tolerant quantum computation.
%We propose that the superior
%computational power of these noise resilient quantum computation
%comes from the non-classical nature of the computation compared to
%a classical nature of noise and will greatly be reduced or even
%be diminished for other noise models which may be realistic.

%Fault-tolerant quantum computation is known to be possible if 
%the noise is ``local'' namely 
%a tensor product of operators acting on individual qubits
%or even if it is $k$-local allowing tensor products of operators acting 
%on non-overlapping ``blocks'' of $k$ qubits, when $k$ is bounded. 
%We show that if ``locality'' is relaxed and  
%we allow the contribution $w_k$ to the noise operator 
%of $k$-local operators 
%to be positive but decay in an arbitrary fast manner as a function of $k$ 
%then fault-tolerant computation is impossible.

%Let $L(k)$ be the class of noise operators 
%that can be described as tensor products 
%of noise operators acting on at most $k$ qubits. 
%Let $L(k)[\epsilon n]$ be the class of such operators where 
%the amount of noise is 
%$\epsilon n$. It is known that if $\epsilon = \epsilon (k)$ 
%is sufficiently small than there are efficient 
%fault-tolerant quantum computation (briefly, FTQC) 
%against noise operators which are in $L(k)[\epsilon]$. 
%In contract, we point out that 
%simple noise operators $T$ is in $\sum L(k)[a_k\epsilon n]$ 
%where $a_k$ is a sequence of positive real numbers 
%which decay in an arbitrary fast way, (say, $a_k = 1/2^{2^k}$) 
%will fail any form of computation.
%(We simply regard certain random operator in $L(k)[a_k\epsilon n]$.)  

We consider stochastic models of 
quantum computation on $n$ qubits subject to
noise operators that are obtained as products of tiny noise operators
acting on a small number of qubits.
%the following form of noise. Let $\epsilon>0$ be a small real
%number and let $\delta >0$,  $\delta << \epsilon$ be an even
%smaller real number. 
%%%(we can think about a situation where $\delta$
%%%tends to 0.)
%The noise is a product $T_1 T_2...T_m$ where each noise operator
%operates on a small bounded number of bits. Each $T_i$ is
%$\delta$-close to the identity and the total amount of noise comes
%to $\epsilon n$. 
We conjecture that 
for realistic random noise operators of this kind 
there will be substantial dependencies between the noise 
on individual qubits 
and,
in addition, we propose that the
dependence structure of the noise acting on individual qubits
will necessarily depend (systematically) on the dependence
structure of the qubits themselves. We point out that the majority function
can repair, in the classical case, some forms of stochastic noise of this kind
and conjecture that this healing power of majority has no quantum analog. 
The main 
hypothesis of this paper is that 
these properties of noise are sufficient 
to reduce quantum computation
to probabilistic classical computation.
%We further conjecture that operators $T$ representing states of
%$n$ qubits in noisy quantum computers are confined to rather small
%spaces of operators: $T$ is approximately $T_1 +T_2$ where $T_1$
%has only entanglements among bounded number of qubits and $T_2$ is
%(up to classic operations) a stochastic noise operator.
Some potentially relevant mathematical issues and problems will be
described. 
Our line of thought appears to be related to that of physicists
Alicki, Horodecki, Horodecki and Horodecki [AHHH].

\end {abstract}

%\newpage

\section {Introduction}
\subsection {Background}

The notion of quantum computation is certainly an exciting intellectual and
scientific development.
Perhaps
the most important result in this field and
certainly a major turning point was
Shor's discovery [S1]
of a polynomial quantum algorithm for factorization.
While some people dismiss the whole idea as a priori too far-fetched and
others even
regard Shor's discovery as an indication that sooner or later a
polynomial classical algorithm for factorization will follow,
it is fair to say that the scientific community
regards the construction  of quantum computers,
which are more powerful than ordinary computers, as a serious possibility.
Whether computationally superior quantum computers 
are possible
%can be built 
is
an exciting puzzle - from an
intellectual, scientific, and technological point of view.

An early critique of quantum computation concerned the matter of noise
which must exist for quantum systems. The possibility of achieving 
fault-tolerant quantum computation (FTQC)
%quantum noise-resilient algorithms 
was demonstrated by a series
of brilliant papers. Shor
showed that quantum error-correction is possible and with
Calderbank [CS] developed this matter further.
Shor [S2] also showed that quantum computation resilient to
polylogarithmically-small noise is possible and
Aharonov and Ben-Or [AB2] and several other groups
(Gottesman, Evslin, Kakade and Preskill; Knill and Laflamme; and Kitaev, see
Aharonov [A1], Preskill [P] and Kitaev [K1])
showed that resilient quantum computation
to a noise that effects a small fraction of qubits is possible.
%These papers demonstrated the feasibility of 
%fault-tolerant quantum computation (FTQC)
In all these papers, it was assumed that
the noise is ``local'' ( a tensor product). In other words, 
the noise operators on
individual qubits (or sometimes several qubits involved in a quantum gate)
%or even small blocks of qubits, 
are independent.
%However, current FTQC methods applies even for much more general
%models of noise.  

The purpose of this paper is to try to find models of noise
that are damaging to current fault-tolerant quantum algorithms
and potentially to quantum computing in general.
%This paper is aiming to attempt attacking  a
%skeptical look on quantum computers primarily based on
%theoretical computer science type of questions.
Our basic point of view is that of theoretical computer science.
The basic complexity-theoretic question is:

\begin {prob}
\label {p:main}
Can quantum computing be reduced to classical (probabilistic)
computing for
models of noise other than those assumed in current
%noise-resilient 
fault-tolerant
algorithms?
\end {prob}

I am thus interested in (even hypothetical) models of noise acting
on a system of $n$ qubits where dependence between the noise
operators acting on individual qubits is permitted. 
%When it comes to 
%analogous scenarios, models
At this stage,
I am mainly trying to get the problem right, and 
consider potentially relevant 
%put in place some
%mathematical infrastructure. 
mathematics.
I pose
some conjectures that are biased against the hypothesis of 
fault-tolerant quantum computing.
%\footnote {Of course, this simply 
%represents the writing style
%of this paper. I would give a reasonable probability that quantum computers 
%will be built and also a reasonable probability 
%that some serious theoretical 
%obstacles will make them infeasible.} 
%%reason that they will not be built
%%a pessimistic
%%view on quantum computers.
%(Some of these, have already been shot down and replaced.)
In the course of this study we will consider 
%(sometimes via questionable analogies) 
some problems and conjectures of independent
interest concerning noise, noise sensitivity, 
Boolean functions, 
%and their
%Fourier transform, 
random walks on groups of 
operators, and error-correction.

It is worth mentioning that already Aharonov and Ben-Or have shown 
%in an
%earlier paper 
that for certain types of noise, e.g. a sufficiently ``strong''
noise that is a tensor product,
a quantum algorithm can be (polynomially)
simulated by a classical one. In these cases the noise is sufficiently
strong to prevent entanglements of more than a logarithmic number of qubits.

Quantum computers works on qubits (say $n$ of them) that are at
each stage in a probabilistic position (state): namely, each of
the $2^n$ strings has some probability which is the (normalized)
absolute value of its (complex) coefficient. These probabilities
are described by a unit vector $U$ in $  {\bf C} ^{2^n}$, and it is
convenient to think about the state of the $n$ qubits as expressing
a unitary operator $S$ acting on an initial state. 
%We can assume
%this initial state gives probability 1 to the the all 0 vector.
The unitary operator $S$ expresses the computation carried out by
the computer starting with the initial state. 
This description (allowing for a measurement at the end 
of the computation) is general enough to describe quantum computers.
The position of the
computer is subject to noise which is usually described by an
operator $T$ involving the $n$ qubits and their environment.
To describe the state of a quantum computer 
subject to noise we need more general objects referred to as 
density matrices. We represent $U$ by a rank one matrix 
$U^{*}\cdot U$ and consider the convex hull of all such matrices.
(A density matrix can thus represent a classical probability distribution
on ``pure'' states.) 
General noise needs to be described by a quantum operation 
which is more general than a unitary operator.
%by a density matrix
%but when talking about general 
%noise operators we should allow also 
%convex combinations (or classical probability distributions) on such 
%basic positions. Those are commonly described by density matrices:

For background on quantum computing, see 
Nielsen and Chuang's book [NC] and also 
Dorit Aharonov's
survey paper [A1] and Kitaev's survey article [K1]. Greg Kuperberg's
emerging book [Ku] is a useful source for the 
mathematics of quantum physics,\footnote {Kuperberg raises the idea that 
quantum physics and the related ``non-commutative'' probability may have 
applications to pure mathematics, similar perhaps to the the role
of the ``probabilistic method'' in various areas of mathematics.} 
and quantum operations in the context of quantum computers.  

For models of
noisy quantum computers it is usually assumed that the probability for 
a ``faulty qubit'' or the ``rate of noise'' is $\epsilon$ 
for some small but not negligible positive
real number $\epsilon$.\footnote{This refer to the rate of noise per qubit 
for one cycle time of the computer.}
I tend to think of the ``amount 
of noise'' of a noise operator $T$ in terms of the Hilbert-Schmidt norm
of $(I-T)$. (A more appropriate norm defines the commonly used ``fidelity''
measure for noise.)  
%total amount of
%noise is $\epsilon n$, 

%The ``amount of noise'' of a a noise operator $T$, $e(T)$  can roughly 
%be described 
%as the expected numbers of faulty qubits. 

%(A qubit can have two types of errors) 
%We consider the action of $T$ for the case that the state of the computer is 
%$o^n$. (Namely all qubits have the value 0.) Then $a(T)$ is 
%the expected number of qubits
%attaining the value '1'.
%For a noise operator $T$, I think of the
%"amount of noise" as the Hilbert-Schmidt norm of $(I-T)$. (I
%realize that there are more appropriate norms.) 

\subsection {The attack}

The best ``attack'' I can see at present is three folded.
%two legs:
\begin {itemize}

\item [a)] Noise operators that deviate a little from the assumption
of being tensor products may kill any form of computations,
%This refer to noise operators 
%where the entanglement between sets of $k$ qubits are not zero 
%but decays arbitrary fast. 

and

\item [a')] 
Noise operators that act infinitesimally on a small number of
qubits may lead to a substantial dependence 
%entanglement 
between the noise
operating on individual qubits.

\item [b)] The dependence of the noise operators on individual qubits
is related (systematically) to the dependence of the qubits themselves.

\item [c)] Devastating stochastic noise considered in a) and a') can be 
healed by ``majority'' in the classic case, but 
cannot be repaired in the quantum case. 

\end {itemize}

Part a') appears to be
similar to a critique proposed by Alicki, Horodecki,
Horodecki, and Horodecki in [AHHH] and their
model 
appears to be related also to part b). 
%Part c) is related to 
%ideas and results by Gottesman.
%%( Part a) seems to be in conflict with claims that FTQC 
%%works for exponential decay of entanglements, however they 

\subsection { Some notations and relevant classes of operators}

I will now describe some classes of operators that will serve us later. 
In particular, we will consider two interesting filtrations 
on the class of all operators. When we talk about ``all operators''
the first %(but unsatisfactory) 
class that comes to mind is the 
class of all unitary operators. It is possible to follow 
most issues raised in this paper having unitary operators in mind,
and, in particular, to consider the definitions here 
as applying to unitary operators. 
The correct class of ``all operators'' is the class of 
quantum operations. 
%(Alternatively, one can think about unitary operators
%acting on the $n$ qubits and 

A quantum operation is a linear map on density matrices
which can be written as
$$    E(\rho) = \sum_k E_k \rho E_k^*,  $$
for some operators $E_1,\dots,E_k$ such that
$$    \sum_k E_k^* E_k = I.$$ 
%The definitions
%in this subsection extend to the class of quantum operations.) 
A different way of thinking about quantum operations 
is to consider just unitary operators but on a larger space --- 
on our original $n$ qubits and their environment.
We can regard the environment to be represented also by some additional 
qubits. These two ways of thinking about quantum operations are known 
to be equivalent and the definitions we will give here apply to both of them.

%For background on quantum computing I used and I can 
%recommend Dorit Aharonov's
%survey paper [A1] "Quantum Computing". Kitaev's survey [K1] is
%also a valuable source and 
% For a noise operator $T$ I think of the
%"amount of noise" as the Hilbert-Schmidt norm of $(I-T)$. (I
%realize that there are more appropriate norms.) 

There are some important classes of operators that we want to consider.

{\bf 1}. $L(k)$ - Operators that are ``$k$-local.'' 
An operator in $L(k)$
can be expressed as tensor products of (arbitrary) 
operators acting on disjoint blocks of qubits
each involving at most $k$ qubits.

Operators in $L(1)$ that act independently on qubits are of 
special importance.

We denote by 
$L(k)[t]$  those noise operators in $L(k)$ where the ``rate of noise'' 
is at most $t$
and we continue to use square brackets to denote an upper bound on the rate of 
noise for other classes of operators as well.

{\bf 2}. $L(k,\delta)$ - Operators that are approximately
``$k$-local.'' They have an $\delta$-approximation by an operator
in $L(k)$.

A class of operators is (uniformly) {\it approximately local} if for
every $\delta>0$ there is $k=k(\delta)$ so that $S \in
L(k,\delta)$.

{\bf Remark:} Current fault-tolerant quantum computation algorithms
resist (even malicious) noise operators in $L(k)$ when $k$ is a
fixed positive integer, provided that the rate of noise is 
sufficiently small. However, classes of approximately local noise operators 
can be very damaging for FTQC.
 
%(And even linear combinations
%of operators in L(k).)
%These noise operators can even depend
%maliciously on the state of the computer. 
%(Namely, a powerful adversary who can read the state of the computer
%without affecting it(!) can decide the noise operator.) 
%The noise can even have
%non-linear ingredients: once the set of error-qubits are determined
%the noise on these qubits can be arbitrary. (We will mention later
%that the current fault-tolerant  algorithms resist even much
%larger classes of noise operators.)

{\bf 3}. $M(\le k)$ - Noise operators $T$ on $n$ qubits that 
represent at most $k$ errors. %(We will make it more precise later.)
To make a formal definition we need the expansion in terms of 
products of Pauli operators. 
Operators $T$, whether they describe the state of the computer or
the noise can be expressed as sums

$$\sum q_v K_v$$
%\noindent
where the $K_v$ is a product of Pauli operators 
and the vector $v$ indicates which Pauli operator operates
on which qubit. We can thus regard $v$ 
as a vector in $\{0,1,2,3\}^n$.
Put $|v|=\{i: v_i \ne 0\}$. We will refer to $K_v$ as a 
multi-Pauli operator of height $|v|$.
Error-correction operators are linear so the
expression of the noise in terms of multi-Pauli operators is important
in understanding error-corrections. The space $$M(\le k)$$ is the 
space of operators that can be described as linear combinations of 
multi-Pauli operators of height $\le k$.
% acting non trivially on at most $k$ qubits.
 
%When we consider the state of the computer and
%for certain classes of noise (that we tentatively call "memoryless")
%it can be assumed that the coefficients
%represent a probability distribution.
%In more complicated cases of noise
%the coefficients are themselves (non-orthogonal) vectors.

We will denote by $w_i$ the overall weight of multi-Pauli
operators of height $i$.
The quantity

$$e(T) = \sum iw_i$$
can be regarded as a measure 
for the ``amount of error.'' 
Similarly, we can think about the quantity
$e_k(T)$, which is the overall weight for all multi-Pauli operators 
acting non-trivially on the $k$th qubit, as a measure of 
the ``rate of error'' for the $k$th qubit. (Those measures have 
the disadvantage of being base-dependent, but they 
can still serve us.) We will denote as $\epsilon$-noise noise operators
(or operations) where the rate of noise for every qubit is at most $\epsilon$.

{\bf 4}. $ILS (\mu)$ - Infinitesimally local stochastic operators.

Operators in $ILS (\mu)$  can be expressed as the product of
random infinitesimally local operators, each acting on a 
small number of qubits,  
according to some
distribution $\mu$. (When considering noise operators we will
discuss later what the limitation for $\mu$ are.) One of the main
points of this paper is to consider noise operators in $ILS(\mu)$
rather than in $L(k)$.

\subsection {Error-correction}

Quantum error-correction is in the heart of FTQC, although FTQC represents 
a long and difficult way beyond 
error-correction.\footnote {However, note that implementing error-correction
requires FTQC}  Fault-tolerant quantum 
computing is thus one of the recent splendid meeting points of 
the theory of error-correcting codes and the theory of computation.
An attack
on FTQC is essentially an attack on the feasibility of 
quantum error-correction.
Computation makes error-correction harder because it tends to amplify errors
and create dependencies among them. A critique of error-correction in the 
context of quantum computers is relevant to general quantum 
error-correction, since quantum computers appear to be an appropriate 
model for any physical device that creates entanglements.
One important insight concerning fault-tolerant computation 
is that it requires a large amount of parallelism. (An early result
of Aharonov and Ben-Or asserts that sequential noise-resilient 
quantum computing is not possible.)

It is useful to keep in mind a certain schematic process
of quantum error-correction and we will briefly describe such 
a process.
In this process (see, e.g., [NC]) an error-correcting code is used so that 
$n'$ qubits are encoded using a larger number of $n$ 
qubits allowing error-correction. The first step of ``detection'' 
is to measure the noise. 
The noise is stochastic but measuring it determines it and this is done
without measuring (and thus without affecting) the signal itself. 
After this step the ``syndrome'' of the noise --- 
a certain multi-Pauli operators on the qubits --- is determined. 
The second step of ``correction'' corrects the errors by applying the 
reverse operations to the faulty qubits. 
This works well 
if the noise is in the correction capabilities of the code. 
%or by throwing away the 
%corrupt qubits and replacing them by correct ones on fresh new qubits. 
%(Sometimes, when we assume 
%a classical computer running aside the quantum 
%one we do not even to carry out the correction.  

\subsection {The paper } 

Here is a brief description of the structure
of this paper. Section 2 considers our basic model of noise based
on infinitesimally local operations and the effect of ``malicious''
noise of this form. Section 3 considers random noise operators. 
We first consider scenarios that deviate a little from the assumption of
locality which may already lead to devastating forms of noise.
%Let $\epsilon_1,\epsilon_2,\dots$ be 
%a sequence of positive real numbers which may decay arbitrary fast.
%We point out that certain random noise operators in 
%$$L(1) [\epsilon n] + L(2) [\epsilon_2 n] +L(3) 
%[\epsilon_3 n]+\cdots, $$
%may fail any form of computation. 
Next we consider
infinitesimally
local stochastic noise operators.
%Understanding such operators appears to be
%the most immediate task for future work. 
Such operators, which seem quite realistic, 
%appears to 
may pose a difficulty to
current noise-resilient algorithms, but it is difficult to see 
them 
reducing quantum computing to classical computing unless 
the noise ``kills'' all forms of computation. 
Sections 4 and 5 give two suggestions  on how to overcome this difficulty.
Section 4
proposes to consider models of stochastic infinitesimally local
noise that are in relation to the state of the computer. 
A systematic dependence between the noise and the state of the computer
%may impose non linear relations (in the form of inequalities) 
%This
%additional ingredient 
has the potential of reducing quantum
computation to a classical one. It also has the potential of
giving a coherent noise model which puts noisy quantum computers
and noiseless classic (digital) computers under the same roof.
%The simplest hypothetical rule
Section 5 studies aspects of the majority function. 
The majority function can potentially repair, in the classic case, 
forms of noise analogous to those considered in Section 3.
This may 
be relevant to the type of spontaneous error-correction taking place 
in digital 
computers (in the microscopic level), and possibly have no quantum 
analog. 
This observation provides an explanation for why classical 
computation may prevail for the type of noise considered in Section 3. 
We discuss also the dichotomy between noise sensitivity and noise stability of
Boolean functions and give an example of noise sensitivity
of a certain model of elections, where it seems that the noise
``conspires'' to spoil the outcome. 
%The third is the theory of 
%Boolean functions and their Fourier expansions that
%can serve as analogs for noise operators and their expansion in terms
%of Pauli operators. The fourth issue is that of error-correction.
In Section 6 we will look at FTQC from another angle.
Assuming that FTQC fails and specifically that the 
noise model proposed in Section 4 is damaging, we
try to understand potential restrictions on the states of
the qubits of the computer. 
We propose that outside the
``neighborhood'' of classical physics, %($L(k)$ for a fixed $k$) 
noise
is essentially all that is left.
%In section VII we consider a
%certain (non-linear) measure of entanglement, and a certain
%related non-linear form of decoherence.
Following a summary of the main problems in Section 7, 
Section 8 concludes. 
Sections 9---11 elaborate on
some of the issues discussed in the main body of the
paper.

\section { Malicious noise}

The known noise-resilient quantum algorithms apply when the
noise is small, has the form of a tensor product, and is malicious
(supplied by an adversary in order to foil the computation). 
%In
%these cases the noise operator can even depend on the "signal"
%itself, namely the state of the qubits in the computer.

There are two models of computation we can consider.
The ``pure'' model consists of just the quantum computer. 
In another model of computation, referred to as ``mixed'' or ``hybrid,'' 
in addition to the quantum computer we have a noiseless classic computer
running aside.
In such a model the quantum computer can be 
at the very least
a source of random bits for the classic computer.
It follows from results concerning randomization in computation
(e.g., Cohen and Wigderson (1989)) that a small noise (of any kind)
will still allow for randomized (classic) computation
based on the $n$ random qubits supplied by the quantum computer.

%When we consider a "pure" model of quantum
%computers (without the digital computer running aside),
%noise of the form described in Conjectures 2.1 and 2.2 can
%potentially harm all form of computation.

It is known that if the noise is in $L(1)[ \epsilon]$ and 
$\epsilon$ is sufficiently small then FTQC and error-correction 
are possible. A basic ingredient in the proof is the fact that 
when expanded in terms of multi-Pauli operators the overall contributions
of multi-Pauli operators of height $k \ge \epsilon n$   
(namely, those which act non-trivially on  
$k \ge \epsilon n$ qubits)
decay exponentially with $k$. In addition, the overall contribution of 
products of Pauli operators that act non trivially on smaller 
sets of qubits that cause harm to the error-correcting code is 
also negligible. It is known and can be proved along similar lines
that if $k$ is bounded then 
$\epsilon=\epsilon (k)>0$ can be found
so that error-correction applies for operators in $L(k)[\epsilon]$.
We start this section  with the following problem:

\begin {prob}
What is the largest growth rate of $k=k(n)$ so that 
for some constant $\epsilon >0$ 
quantum error-correction (and FTQC) applies to arbitrary noise operators
in $L(k(n))[\epsilon]$ acting on $n$ qubits? 
\end {prob}

We will now describe our basic model of noise.

{\bf (2.1)} The noise operator T is obtained by successive applications
of noise operators $T_i$, $i=1,2,\dots,m$ where $T_i$ is $\delta$-close to the
identity. Each operator $T_i$ operates on a bounded small number
of qubits. (We can either consider operators acting 
on a small number of qubits
and their environment, or consider quantum operations acting on 
density matrices that correspond to these qubits.)
%(or a
%bounded number of qubits) and their environment. 
The total amount of noise is $\epsilon n$. 

It is important to note that we allow
``cancellation,'' namely, the amount of noise for 
$T$ is a sub-linear function in terms of the amount of noise 
of the individual $T_i$'s. Such a ``cancellation'' 
can be expected in the stochastic models that 
we consider in the next section. 
Without cancellation, when the rate of noise for $T$ is 
simply the sum of the rates of noise for the $T_i$'s it can be shown that 
up to an exponentially small error 
$T \in M(\le \epsilon' n)$, for every $\epsilon' >\epsilon.$ .   

\begin {conj}
\label {c:2.2}
Malicious noise of this form kills all forms of 
computation in the pure model
and reduces quantum computation to classical (probabilistic) computation
in the mixed model.
\end {conj}

{\bf Remark:} 
There are several possible interpretations for Conjecture 2.2.
If we allow $m$ to be exponential in $n$, the noise 
operator can approximate 
any unitary operator. This appears not to enable any form 
of computation in the 
pure model and to leave us with randomized 
classical algorithms in the 
mixed model. (But it is possible that bounded depth computation 
will prevail.)
It is more interesting to 
consider the case where $m$ is 
polynomial in $n$. I would expect a malicious $\epsilon$-noise  
to be able to kill computation (or at least to reduce it to 
bounded-depth 
computation,) even if generated by a polynomial-size 
polylogarithmic-depth circuit.

%One can argue that qubits which are not entangled to start with will
%not become entangled.

%\begin {conj}
%Conjectures 2.2 continues to hold even if we assume that the
%noise operators $T_i$ do not affect triples of (or a bounded set
%of) qubits which are not entangled to start with.
%\end {conj}

\begin {conj}
%(1) 
Conjecture 2.2  continues to hold even if we insist on the
resulting noise operators to be uniformly approximately local.

%(2) Conjectures 2.2 continues to hold even if we assume that the
%noise operators $T_i$ do not affect sets of 
%qubits which are not entangled to start with.
\end {conj}

%The main point of these conjectures is that a substantial dependence
%between the noise operators on individual qubits
%can be created from repeated applications of
%small noise operators each affecting a small number of qubits.

We also conjecture that malicious noise can be used to
decay ``high order'' entanglements:

\begin {prob} 
Find a malicious infinitesimally local, approximately local,
noise that forces the state of the quantum computer to be
uniformly approximately local.
\end {prob}

{\bf Remark:} 
A work of Tsirelson and Vershik [TV] (and also a work by Benjamini,
Kalai, and Schramm [BKS])
suggests that by repeated application of a noise operator
to three qubits
a substantial dependence  between the noise operators
on individual qubits may result 
(this appears to move us away from the tensor
product assumption used in error-correction and 
fault-tolerant algorithms).
Tsirelson and Vershik showed
that in a recursive ternary tree, aggregation of every generic
function from the leaves to the root will have such an effect. 

\section {Oblivious random noise} 
%- hunting for a power law decay} 
%(stochastic infinitesimally local
%noise)
%}
\label {s:stoc}

\subsection {Arbitrary random noise}
\label {s:3.1}
As we mentioned in the Introduction the presentation of a noise operator 
in terms of 
products of Pauli operators and, even more, in terms of 
the filtration $M(\le k)$ 
of the space of noise operators is important for the issue of 
error-correction. 

We will consider in this section models of noise that are 
invariant under permutations of the $n$ qubits. 
Under this assumption error-correction and current FTQC prevail if  
the the noise operator ``approximately'' (up to an exponentially 
small error) 
belongs to $M(\le k)$ where $k=\epsilon n$, for some 
specific small $\epsilon >0$. 
When it comes to error-correction the 
explanation is easy. Error-correcting 
codes fail only when the syndrome consists of a relatively large fraction
of all qubits or (in case, say, of the concatenation code) 
of rare smaller 
``bad'' subsets of qubits. By the assumption of invariance the 
correction will rarely fail.
Random models of noise that are invariant under permutations 
of qubits will be damaging only if in their multi-Pauli expansion
a large amount of weight on high multi-Pauli operators  
is present.

We will start with some 
basic observations and questions. When we think about the expansion 
of a random operator in terms of the basis of multi-Pauli operators
we can expect that most of the weight of the coefficients will be on 
multi-Pauli operators 
of heights  around $3/4n$. 
The reason is simply because most multi-Pauli operators
 are of these heights. (This can be 
regarded as a ``concentration of measure'' argument.)
If we assume the noise rate is $\epsilon$ (and here 
there may be some delicate points on how to measure the amount of noise),  
still we can expect the weight of multi-Pauli operators 
on at least $0.74n$ qubits to be large (over, $\epsilon/2$, say.)
Such a noise is quite far from our intuitive way of thinking about 
noise of rate $\epsilon$ as it represents events that 
with substantial probability
corrupt of a majority of all qubits.\footnote{Imre Barany proposed
the following analysis which may demonstrate the effect of dependence. 
Wars between two neighboring countries erupted in Europe 
from time to time.
Towards the end of the 19th century dependence caused by 
a large amount of treaties between 
countries led to %substantial dependence: The effect was 
a long period of peace --- followed by a world war, involving 
almost all European countries.} 
Arbitrary random noise operators acting on all qubits 
is probably not something that we should worry about, but we ask if 
similar properties of noise can be found in more realistic scenarios. 

\begin {prob}
\label {p:blocks}
1. Let $T=T_1 \cdot T_2\cdots T_m$ where $T_i$ are random operators in 
$L(2)$ (the partitions to blocks are also random). Suppose that 
each $T_i$ represents a rate of noise $\delta$,
$T$ represents an expected rate of noise $\epsilon$, and 
$m$ is chosen accordingly. What will the expansion of $T$ 
in terms of multi-Pauli operators look like? 
(We may think of the 
case $m$ is logarithmic in $n$.)

2. The same question as in part 1) except this time 
suppose that $T_i$ is a random operator in $L(k)$ and that 
$k$ grows to infinity 
very slowly with $n$. (Again, we may think of the 
case where $m$ is logarithmic in $n$.)
\end {prob}

Concerning  part 2 of Problem \ref {p:blocks} the following heuristic argument
suggest that 
(as $n$ tends to infinity) indeed the multi-Pauli expansion of $T$ will 
be heavily concentrated for heights larger than  $0.74 n$:
Consider a stochastic operation $T_i$ in 
$L(k)[\epsilon]$ when $k$ grows slowly to infinity with $n$ and 
$\epsilon$ can be very small (and even tend to 0 with $n$).
Thus, $T$ is a tensor product of operators on non-overlapping blocks of size
at most $k$. From the observation concerning arbitrary random operators it 
follows that 
qubits in the same block will be very correlated. Now, taking products of
several such operators with random partitions to blocks may have the 
effect of making all qubits highly correlated. 

Finally, consider the following 
model of noise:

{\bf (3.1)} The noise operator consists
of taking products of random operators in $L(k)$ 
where $k$ itself is a random variable 
whose distribution $D(k)$ is positive and decay to zero with $k$. 

It appears that the 
computational power under such a model is that of a bounded-depth computation 
where the bound on the depth depends on the decay behavior of $D(k)$.
(Compare, however, Section 5.1 which suggests that in some 
scenarios classical computation 
may prevail.)

\subsection {Infinitesimally local random noise}
Let us return now to the model 
of random products of tiny operators acting on a bounded number of qubits.
%But are Conjectures 2.2 and 2.3, even if true, (and even if the
%model of tiny noise operators operating on a small number of qubits
%realistic) damaging?

Letting an adversary choose the local noise operator in Conjecture 2.2 
is too harsh. Suppose that the tiny (or infinitesimally) local
operators are chosen uniformly at random according to some
distribution $\mu$. We will now discuss the following conjecture:

\begin {conj} 
\label {c:stoch}
Quantum computation subject to (realistic)
random noise of the form described in relation (2.1)  above when
the qubits on which the noise is applied are chosen uniformly
at random is (polynomially) reducible to classical (probabilistic)
computation.
\end {conj}

%As we will see in this section Conjecture \ref {c:stoch} 
%appears to be false

%but stochastic noise of this kind can still be potentially damaging.
%We cannot present a case where current QFTC fails and there 
%is a good defense even in cases where current FTQC

%%It appears that the property
%%of the noise being a tensor product is not crucial as long as the
%%dependence  between noise operators on individual qubits
%%is not too strong and is not related to (say) the structure of the code
%%used in the Aharonov-Ben-Or algorithm.

Following the discussion above, the 
crucial question is thus whether we can ignore the contribution
of very large products of nontrivial Pauli operators acting 
on very large sets of qubits (say more
than 74\%)? Is it the case that for certain choices of the
distribution $\mu$ the weight of multi-Pauli operators acting nontrivially 
on very large
subsets of qubits will be bounded away from zero? or perhaps be 
polynomially small but not negligible?

%(Namely, $\sum w_i: i>0.73$ would be of size $n^{-\beta}$ for some
%$\beta$. 

%I expect that
%this will occur when a substantial amount of ``cancellation'' for the
%tiny (or infinitesimal) noise operators takes place.
%This can be damaging, and may fail the Aharonov-Ben-Or
%algorithm.

%%%The following problem looks doable and among all problems in the paper
%%%it is perhaps the first thing to try to do.

\begin {prob}
\label {p:3.2} Show that for the models described in Conjecture
\ref {c:stoch} for an appropriate choice of the distribution $\mu$:
\begin {itemize}

\item[(a)]The noise operator is approximately local:
For every $k$ there is $\epsilon = \epsilon (k)$ so that $T$ can
be approximated by an operator $T_1$ in $L(k)$.

\item[(b)]
The contributions of multi-Pauli operators acting non-trivially 
on $k$ qubits is bounded away from zero (say, when $k \le 0.7n$).

{\rm or} 

\item[(b')]
The contributions of multi-Pauli operators  
acting non-trivially on $k$ qubits decay
as a power of $k$, $k^{-\beta}$ , $\beta >0$, when $1 \le k \le
n$.

\end {itemize}
\end {prob}

%{\bf Remarks:} 
%In view of Section \ref {s:3.1} 
Perhaps the best shot for a distribution $\mu$ for this problem will be if 
$\mu$ allows random operators on $k$ qubits with positive 
probability that may even decay very fast with $k$. When $k$ is large, 
a random 
noise operator on a block of $k$ qubits creates a large correlation between
these qubits being faulty. The model of applying successively 
noise operators on small blocks of qubits is very close (perhaps even 
identical) to the model of noise (3.1) 
of the previous subsection.   
%It is possible
%that when $\mu$ represents operators acting only on at most $k$ qubits
%the noise operator will behave like operators in $L(k)$ and another 
%possibility is that interesting phenomena occurs already for $k=2$.

An interesting case of Problem \ref {p:3.2} is that
of random products of tiny unitary operators. 
Let $W$ be a class of unitary operators acting on (at most) 
pairs of qubits, suppose
that $W$ is closed under inversion and suppose that 
each operator in $W$ represents a noise $\delta$.  
The simplest case to consider is when $W$ consists of 
two tiny rotations operating on a single qubit and two tiny rotations 
in the direction of CNOT operating on two qubits.
%(We can also consider the case that $W$ is a universal set for 
%quantum computing.) 
Let $G$ be a graph on $n$ vertices (the qubits of the computer).
We can assume that $G$ is the complete graph with loops.
%$T_1,T_2,\cdots,T_s$ represent (for simplicity) 
%unitay noise operators each representing 
%an amount $\delta $ of noise. (For unitary noise this 
%simply means that if the signal is $o^n$
%with probability 1 to start with,  the expected number of non zero qubits 
%afer the noise is $\delta$.) 

Consider a random product $T=T_1 T_2 \cdots T_m$ of length $m$ of 
such operators. Thus each $T_i$ is a random operator from $W$ applied 
on the qubits of 
a random edge of $G$. (This is a random product with $e(G)|W|$ generators.)
Let $E(m)$ be the expected amount of noise of $T$.
Let $m$ be chosen such that $E(m)=\epsilon n$ and let 
$T$ be the resulting random operator.

When $\delta$ is large enough there will be essentially no cancellations and 
the behavior will as in the case of local noise operators. Understanding 
this model when $\delta$ is small (or ``intermediate'') is of interest.
%\footnote {Greg Kuperberg proposed an argument suggesting  
%%pointed out 
%%(see Section \ref
%%{s:defense}) 
%that when $\delta$ is very small 
%(or infinitesimally small), although there will be a substantial 
%amount of cancellation 
%%(as in the case of a Euclidean random walk), 
%still the contributions 
%of multi-Pauli operators acting 
%non-trivially on $\epsilon \cdot n$ qubits or more 
%%($\epsilon'$ slightly larger than $\epsilon$) 
%will decay 
%exponentially with $n$. 
%%(The reason is that the contributions of high-order 
%%commutators become negligible. Greg's original argument assumes a bounded
%%number of generators so I am not sure if it applies in our case.)  
%%Therefore, for this case current FTQC may apply. 
%Our best shot for large contributions of high multi-Pauli operators
%remains for the case that $\delta$ 
%is intermediate or possibly critical between the two different behaviors 
%for large and small $\delta$.  
This model looks quite close to the 
Ising model on graphs and its analysis may be feasible.

There are various examples in the literature 
of how local stochastic operations may lead to 
substantial dependencies. Valiant [V] gives an example of how starting with 
random independent bits and performing local stochastic operations
we can reach with high probability the majority function. This result 
suggests that 
starting with noise operators acting independently on $n$ bits 
we can reach, by local stochastic operations  
that preserve the marginal probabilities of bit-errors, a substantial 
amount of dependencies.

% and 
%independently interesting.

%{\bf Remark:} In view of the previous section there can be crucial difference
%between the case that $\mu$ involves 
%only actions on bounded number of qubits 
%or allows sufficiently fast decay with out having an upper bound. 
%In any case, I fine the problem
%of products of tiny operators acting on pairs 
%of qubits a very interesting one. 

{\bf Remarks:}
1. Recall that our principal assumption is the invariance of the noise
model on permutations of qubits. There are various
reasons why this symmetry could be broken (and in a damaging 
way). A primary (hypothetic) such reason that 
we consider in the next section has to do with
the entanglement structure of the ``signal,'' which may be echoed by the noise.
Another (related)
reason for breaking this symmetry has to do with the structure of the
circuit itself and the gates involved in the computation. The probability
distribution on tiny (or infinitesimal) noise operators may depend
on the circuit's structure and the identity of qubits that belong to the same
gate. Still another reason
is related to the hypothetical geometry of the quantum computer.

2. The possibility of a polynomial decay 
(in terms of the projection on 
$M(\ge k)$ ) 
rather than an exponential decay is interesting but 
I am not aware of any %(potentially realistic) 
concrete infinitesimally local 
stochastic noise model that exhibits such  decay. 
Suppose we did find an example of a noise operator for which the
decay of the coefficients in the expansion to  
multi-Pauli operators 
satisfies a power-law decay with the height. How damaging would this be? 
Dorit Aharonov suggested
a defense against such power-law decay for the noise 
for the mixed model of quantum computers: 
For every $T>1$, an algorithm on $n$ qubits
%with running time, say $n^2$, 
can be replaced in the mixed model 
by an algorithm on $n^T$ qubits with the same running time.
(This is not known and perhaps even false in the pure model 
of quantum computers.) If the number of qubits is sufficiently 
large compared to the running time current FTQC will prevail.
%Thus, it
%appears that Conjecture 3.1 is false. 
This shows that a polynomial-decay in terms of expansion to 
multi-Pauli operators of stochastic 
noise operators does not harm the computational 
power of quantum computers. (But it can be practically problematic.)

3. It can be argued that the ``tiny'' operators used in our model
may not act on qubits which are far apart according to a
hypothetical geometry of the quantum computer, or that they should 
respect the architecture of the computer. 
Similarly, it can be argued that the blocks considered in 
the problems of Section \ref {s:3.1} should also respect a 
hypothetical geometry of the computer.
I would expect that
under reasonable restrictions of this kind matters 
will not change. In any case, the graph $G$ considered above
may reflect the geometry or architecture of the computer.

\subsection {Modeling noisy computation}

\label {s:model}

We conclude this section by noting that
from the point of view of complexity theory (where it is natural 
to consider a
``pure'' quantum computer)
it appears that none of the variations of the
basic model of stochastic infinitesimally local 
noise considered in this section
have the potential to reduce quantum computation to classical computation
without killing all forms of computation (beyond bounded-depth 
computation).\footnote {In Section 5.1 we suggest that in 
some scenarios with unbiased stochastic errors 
effecting large percentage of qubits, classical computation 
may prevail.}

Notice that we have a difficulty with the model. Unlike quantum
computation which is a robust (and quite wonderful) model of
computation, giving what appears to be a clear complexity class,
noisy quantum computation is problematic. It appears to be a
difficult task to base a complexity-theoretic 
attack  on quantum
computation on a noise model which affects classical computation as
badly as quantum computation and certainly if the noise kills all
forms of computation.

It is hard to base a model of computation on a statement like:
``Quantum computers will have a substantial error rate of at least
$10^{-4}$ ... unless they happen to be ordinary computers, in
which case they will be essentially noise-free''. There appears to
be a basic difficulty in modeling the noise of quantum computers,
which includes ordinary digital computers as a special 
case. 

In the most abstract setting of finding a unified 
noise model for noisy computers with $n$ {\it logical} bits, the hypothesis 
of fault-tolerant quantum computing indeed offers such a model: 
The model of noiseless computation. It would be interesting to describe 
an alternative model with non-zero noise
in the non-classical case, which is consistent with the laws of 
quantum physics. (Of course, in such a model the hypothesis 
of fault-tolerant quantum computers fails.) 

Less abstractly, for studying noise models 
for computers with $n$ physical qubits, a unified 
model for (noiseless) classical and (noisy) quantum computers 
still makes sense and seems necessary for finding scenarios 
where noisy quantum computation reduces to classical probabilistic
computation. 

In any case, 
%as long as we have 
in a reality of sharply different models of noise for 
digital and quantum computers (even in terms of the constants involved),
we cannot dismiss claims that noisy quantum 
computers will not be able to perform any kind of 
computation 
%(like those in 
%Alicki 
%[Al, ALZ]) 
just on the grounds that 
classical computation and classical error-correction 
do exist. But, on the other hand, 
it will be hard to accept any such 
claim against quantum computers 
as completely satisfying.\footnote {Kuperberg mentioned computation 
processes in biology (say, the brain) as examples 
of noisy computation, where the 
model of noise might be closer to the case at hand. Indeed, these 
computations exhibit a substantial amount of parallelism in 
according with insights of fault-tolerance. They also appear to 
represent ``small depth'' computation. I do not know if independence
is a reasonable assumption for noise models in such systems. (I would not 
expect so.)}

%but the case for it is weaker. (We
%come back to this point in the next section.)

\section { Random noise that is neither 
%a tensor product, nor
malicious, nor oblivious, but rather related to %the entanglement of
the signal.}

Noise operators, like all operators in quantum 
physics are linear. Is it possible, though, 
that noise operators satisfy systematic 
non-linear inequalities? Before jumping to a fierce ``no''
note that the starting point of FTQC, the fact that 
quantum computers, unlike digital computers,  are subject to a substantial 
amount of noise, is, at least on the face of it,  
an example of such a non-linear inequality.

An additional attack 
on quantum computers 
suggests that dependencies that are expressed already by rather low multi-Pauli 
operators can already cause problems.
It goes vaguely as follows (we will try to make it 
more explicit later on):

\begin {conj} Realistic stochastic models of noise (based
on tiny noise operators of the kind we considered above) will create
dependence between the noise operators among qubits, which itself is
associated to the dependence structure of qubits yielded by the
quantum computation. In particular, there will be a damaging 
dependence on the block structure of an error-correcting code
used in the current noise-resilient computation. Moreover, this
kind of noise suffices to reduce quantum computing to classical
randomized computing.
\end {conj}

Conjecture 4.1 follows a simple logic of ``reverse engineering,'' i.e.,  
trying to understand
how fault-tolerant algorithms can (badly) fail. It would be important
for this purpose that the dependence between the noise and the signal
apply to high terms in their
expressions in terms of multi-Pauli operators, or, even more directly,
the sets of qubits with large coefficients in the expansion
should be ``badly located'' as far as the error-correction is concerned.

{\bf Remarks:} 1. This claim of a relation between the dependence
structure of the signal and the dependence structure of the noise
may look strange, and it is not a priori clear also why such 
dependence may be damaging. Compare, however, the elections example
in Section \ref {s:election}.

2. It seems suspicious that the stochastic model for noise
may depend on the ``signal'' (the state of the $n$ qubits). 
If the linear operator describing the
noise depends on the signal itself then the noise may depend
non-linearly on the signal. This looks non-kosher and has indeed drawn
criticism. 
However,
in models of quantum computation, in order  to achieve each desirable
distribution among the qubits we need
a different physical device (say circuit) and the noise can very well
depend on
this device. So it appears that the kind of dependence 
we consider is
quite expected and a non-linear (stochastic) 
dependence of the noise on the signal
does not contradict the fact that the operators 
describing noise are linear operators.
To make the argument clearer the reader is referred
to the simple example in Section \ref {s:exa}.

3. A dependence between the
entanglement of the noise operators and the entanglement of the qubits
themselves appears to be related to an argument by
R. Alicki, R. Horodecki, M. Horodecki, and P. Horodecki [AHHH].
According to their argument the neighborhoods of the
qubits will echo the entanglements between the qubits. This may lead
to the type of dependence we propose between the dependence 
structure of the noise and that of the signal. 
%In this case the argument strongly
%depends on the noise having ``memory''.
%%\footnote {I do not know how 
%%to translate the property of the noise as having ``memory'' 
%%to mathematics.}.

4. A serious  critique already raised against the argument of
Alicki, Horodecki, Horodecki, and Horodecki is that it is not clear
why the type of noise they consider will ``conspire'' against the computation.
Parts of the rest of the paper can be regarded as an attempt to understand
the implications of such a ``conspiracy'' which may suggest also why
such a ``conspiracy''
might be possible. The most appealing answer I can think of
is that the dependence
of the noise on the signal is systematic and is expressed by  
%Its effect is of disentangling
%the entanglements, or even forms of classical probability dependence,
%and it thus leads to 
non-linear (as a function of the signal)
relations (inequalities) for the decoherence 
which are 
%can be 
damaging. 
%(Indeed, if such a phenomenon
%accounts for some ``elasticity'' in entanglement formation, ``memory'' seems
%a necessary ingredient.) 
%%I am not sure if such a relation can occur for
%%noise without memory.)

5. 
%Having a non linear form of decoherence --as a function of the signal--
%appears to be consistent with quantum physics.
If there is a systematic form for the dependence of the noise on the signal
(which lead to systematic non-linear relations), we can ask, what is its
mathematical nature.
%If the linear operator representing the noise depends linearly on the signal
%(and we can always consider the linear approximation) this will lead to
%a quadratic form of decoherence. So exploring quadratic
%forms of decoherence can be a good place to start.
%It is certainly  for physicists to tell what type of
%decoherence of this form is reasonable (if at all),
%say, based on the model of [AHHH].
From the complexity-theoretic point of view it 
may present us with an opportunity
to address the problem of finding a model of noisy quantum computation
that is consistent with classical computers
being noiseless. (The dependence of the noise operator 
on the state of the computer is likely to be expressed 
by a differential relation (probably inequalities rather than equations)
describing the dependence of the noise on the state in an infinitesimally 
earlier time. 
%\footnote {{\bf Problem:} Propose a mathematical 
%description for such a differential equation.})

\begin {prob} 
%Propose a non-linear (quadratic?) mathematical
%model for decoherence. More precisely, 
Propose non-linear %(quadratic?) 
inequalities satisfied (necessarily) by (linear) decoherence operators for 
quantum computers.
\end {prob}

It would be nice to have a description not affecting a situation
when the $n$ qubits are independent, and, more generally, reducing
probability dependence (or covariance) between pairs of qubits. 
Some suggestions for non linear inequalities for decoherence
can be found in Section \ref {s:decoherence}. 
%Desirably, the abstract mathematical forms can 
%also be described in a way which does 
%not depend on the
%identity of the qubits, namely,  the sub-gadgets of the physical
%device that are considered the ``logical'' qubits of the computer.
Of course, it would also be needed to relate such inequalities
to infinitesimally local models of noise 
when we allow the infinitesimally local operators acting on a few qubits 
to (stochastically) depend on the state of these qubits.

6. Perhaps the simplest explanation of why quantum computers are 
intrinsically noisy that offers simple non-linear inequalities 
for the (linear) decoherence operators is that correlations are collapsing. 
Can it be that in all 
quantum systems (and perhaps also in classical physical systems) 
correlations between qubits and especially 
correlations between many qubits are fading away? 
As pointed out by Robert Alicki this type of proposed behavior   
seems related to ``Onsager regression theorem'' 
in (classical) statistical physics.

%7. The only way I see to interpret the statement that quantum computers, 
%in contrast to digital computers, are noisy, which was the starting 
%point of fault-tolerant quantum computing 
%is as describing non-linear inequalities for decoherence.

7. We can try to model dependence of the noise on the gates, 
hypothetical relations between the noise and the signal,
and hypothetical ``elastic'' properties of forming 
entanglements,
in a more combinatorial way by adjusting the 
random walk model 
of the previous section.
For example, let the random walk noise 
described in the previous section run 
in parallel to the actual computation carried out by a quantum computer
and add to our stochastic oblivious noise 
after every operation of the computer 
% stochastic quantum computer of Section 3
%a stochastic infinitesimal local noise 
%with fixed distribution $\mu$ of Section 3 
additional random generators (say with probability decaying in time) 
that $\delta$-reverse the 
operation taken by the computer. 
%of any gate some time after it was carried out. 
%%We can also consider 
%%running a quantum circuit with the property that the operation in each gate 
%%has a small chance to be reversed at subsequent cycle 
%%times (rather than immediately).} 

8. Michael Ben-Or offered an ingenious (yet, incomplete) 
argument (related to an argument by Preskill and Shor) to the effect that
a quantum computer (with a classical computer running beside keeping track
of some of the noise), may run so that the state of qubits will be
``completely random.'' 
%(Michael pointed out that a related argument appeared
%in a paper by Preskill and Shor.)
%This may give a defense against a damaging relation
%between the dependencies of
%the noise and the signal. So far it appears (to both of us) that this argument
%is not complete. It looks (to me) that the entanglement structure
%must be ``left'' in the quantum computer.

\section {Stability, sensitivity and the majority function}
\label {s:sensitivity}

\subsection {Merits of the majority function} 

This section's three part are all related to the majority function.
Given an odd integer $n$, the majority function 
$f(x_1,x_2,\dots,x_n)$ is a Boolean function on $n$ Boolean variables
defined by: $f(x_1,x_2,\dots,x_n)=1$ if $x_1+x_2+\cdot+x_n>n/2$, 
and $f(x_1,x_2,\dots,x_n)=0$ otherwise.  
This subsection gives another mathematical suggestion on 
how to reconcile the possibility of fault-tolerance
in the classical case with the possibility of fault-tolerance being 
impossible in the quantum case. 
The idea is that in the classical case, devastating behavior expressed by 
random 49\% (say,) of bits being harmed (in an unbiased way), 
can still be repaired
by the majority function, 
and that this healing power of majority, which may be relevant to 
modeling digital computers in the microscopic 
level, has no quantum counterpart.
The prominent role of 
the majority function for classic fault-tolerance 
%stability to noise 
and the difficulty in 
realizing ``majority'' in various settings 
regarding quantum computers is mentioned e.g. by Gottesman in  
[G] as an important distinction between quantum error-correction and 
classical one. 
The idea that ``majority'' is essentially 
a classic notion and, as some natural extensions of majority to the 
context of quantum states are non-linear, 
cannot be extended to the quantum setting can be found in several 
other places. 
However, I am not aware of a useful formalization of 
this idea in the literature. 
(The majority function is used, in fact, in various quantum 
error-correcting codes.) 

As we mentioned in Section \ref {s:model} one difficulty in 
various attacks on quantum computers, and, in particular, an attack 
based on models like those 
considered in Section \ref {s:stoc}, is that these attacks continue to apply
for digital computing, and are especially relevant when we consider 
digital computers on the microscopic level. 

Current schematic descriptions of digital computers on the microscopic level
are based on each logical bit described by the majority function of a huge 
number of ``physical'' microscopic ``bits''.\footnote{Describing a 
detailed mathematical model 
of digital computers in terms of the 
microscopic representations of logical bits,
including a description of the noise (and gates), will be of interest.} 

It is important to note that the majority function is 
immune against random unbiased 
errors which come very close to effecting 50\% of all 
bits --- like the kind of errors
considered in Section \ref {s:3.1}, but in the classical setting. 
%(We can expect that unbiased stochastic noise will be most damaging
%and that biased noise will have a similar behavior to independent noise.) 
%%I do not know 
%%if this can be expected by mechanisms based 
%%on more subtle error-correction ---
%%even in the classical case. 
%%Of course, 
If the type of noise considered in Section 3 is realistic for digital computers
described in the microscopic level majority-based self-error-correction 
can still prevail. 

On the other hand, I do not know if this is possible
in the quantum case (where 50\% should be replaced by 75\%). Non-linear 
majority-like (or rather ``plurality'') functions on quantum 
states will correct random unbiased errors
effecting almost 75\% qubits but I suspect this cannot be achieved by 
linear error-correction. (Here by ``plurality'' I refer to a function 
that given $N$ states
outputs the state that appeared the largest number of times.)
%As a matter of fact a negative result may well 
%already known by recent results by Gottesman 
%and others.

\begin {prob} (1) Demonstrate that majority-based error-correction 
in which a logical bit is represented by the majority value of a 
huge number of physical bits can repair (classical analogs) of 
stochastic noise considered in Section 3.

(2) Find an argument for showing that this is impossible in the quantum case.
\end {prob}

{\bf Remarks:}

1) It may well be the case that such an argument can be found in the 
existing literature and may be related to the issue of ``optimal cloning''. 
Daniel Gottesman referred me to [BDEFMS] for a related result.
(But for our purposes a standard no-cloning argument may suffice.)

2) A useful way to think both about the classical and quantum case together 
is to apply on  98\% (say) of the 
bits (or qubits) a random unbiased uniform rotation.
For ordinary Boolean bits we can expect 49\% of the bits to be harmed.

3) An exciting direction in the quantum computers
endeavor is the 
constructions (theoretical, so far) of self-error-correcting physical devices. 
Kitaev 
found [K2, K3] an error-correcting scheme based on 2-dimensional
topology which can be regarded as
the starting point of a whole new physical model for quantum computers
referred to as topological quantum computers, see, e.g., [FKLP].
%Nick Read pointed out that any pessimistic result concerning
%quantum computers is not complete unless
%it includes the topological quantum computer model.
The idea
of topological quantum computation is to embed the error-correction
in the physical device, in an analogous way, perhaps, to what make ordinary
digital computers (essentially) error-free.

\subsection {Stability and sensitivity}

(Noise) sensitivity and (noise) stability is a setting where, to
get an advantage over classical (probabilistic) computation, a
substantial amount of ``dependence'' is needed, which also implies a
substantial sensitivity to noise. It is a scenario where there is a
dichotomy between the weighted majority functions (which are
stable) and functions asymptotically orthogonal to them (which are
sensitive). This appears to be related (and perhaps suggests a way
to formalize) an  assertion that is often made that in order to
get an advantage over classical probabilistic computers a
``substantial amount'' of entanglement is required. 
%It may also be related 
%to the informal claims that the majority function is crucial in classical
%fault-tolerance and in classical situations of 
%spontaneous error-correction, and 
%its absence in the quantum setting makes these matters harder.
%For more details
%see Section \ref {s:sensitivity}.
The notion of (noise) stability and (noise) sensitivity
was introduced by Benjamini, Kalai and Schramm [BKS] and
was further developed by various people. It is also closely related to the
work of Tsirelson and Vershik and various works of Tsirelson [VT,T].
(I will just use the term sensitivity and stability since we may want at times
to apply these notions also to ``noise''.) 
This is a setting where in order to have an 
advantage over classic (deterministic) computation sensitivity to noise 
is required and thus it may be relevant for us.

Let us consider a randomized computation which depends on $n$ coin flips
that are independent and unbiased. Suppose that if the answer is NO
the computation gives 0 while if it is YES it gives 1 with a probability 
of at least 1/2.
Suppose that the answer is YES,
the $n$ bits are chosen at random, and the computation yields the outcome $T$.
Suppose next that a fraction $\delta$ of the bits chosen at random are flipped
and the new outcome is $T'$.
Let $g(\delta )$ be a fixed function that tends to 0 when $\delta$ tends to 0.
We say that the computation is (uniformly) stable
if the the correlation between $T$ and $T'$ is at least $1-g(\delta )$.
(This is an asymptotic notion for a class of Boolean functions.)

If the computation is uniformly noise-stable then it can be
simulated by a
polynomial classical algorithm.
(This follows from the basic Fourier description
of stable Boolean functions: a class of Boolean functions is uniformly stable
if most of the L-2 norm of functions in the class is concentrated on a 
bounded number of levels in terms of the Fourier expansion.) 
It is quite possible that $(1/(small~ polylog(n))$-stable or
even $(100/\log (n))$-stable suffice. This and related problems
are described in Section \ref{s:sens.app}
%(Conversely, the assertion that
%a classical randomized algorithm in $poly (n)$ time can be
%transformed to a $(1/poly log (n))$-stable will follow if
%quasi-polynomial random bits suffice for such an algorithm,
%which is far from being known.)

%Stability and sensitivity in this sense are certainly
%notions to keep in mind. 
%This is a setting where in order to have an 
%advantage over classic (deterministic) computation sensitivity to noise 
%is required.
%If we can compute something by Boolean
%functions that are uniformly stable  (probably even to small
%$1/polylog(n))$ noise) then the computation is classic. This may be
%useful. 
%However, since classical randomized algorithms based on
%(even $1/\log (n)$)-stable computations do not capture the full
%strength of randomized algorithm (as far as we know) we probably
%need weaker notions of stability for the purpose of proving
%reductions to classical probabilistic algorithms.

In the world of Boolean functions weighted majority functions are noise-stable
and a sequence of  functions that are asymptotically orthogonal to every
weighted majority function are noise-sensitive [BKS]. 

Finally let me remark that the 
related notions of stable and sensitive stochastic flows 
by Tsirelson (who studied these concepts also 
in the quantum context) 
may be closer to the 
context of quantum computation and noise operators. 
%In the Boolean world
%functions that are closer to the case at hand are functions 
%from $\{0,1\}^n$ to $\{0,1\}^n$ (and more generally 
%from $\{0,1\}^n$ to $\{0,1\}^m$), and indeed extending 
%various aspects of Boolean functions in this direction is on the agenda.

\subsection {An analogy - An example concerning elections}
\label {s:election}

The issue at hand is about noise sensitivity in systems with
probability dependence. 
Following is an example (taken from a paper of mine on social choice [Ka])
which demonstrates some of the issues
that arise when we consider noise sensitivity of Boolean functions
(thought of as elections with $n$ voters)
when the distribution for the input is not a product distribution.
The Boolean function is simple majority but the  voter
behavior is not independent. (Of course, this
is only an analogy to the case at hand.) 

Suppose that the society is divided
into  communities of $b$ voters each. The number of voters is
thus $n=ab$, which we assume is an odd number.

Each voter $i$ receives an independent signal $s_i$, where $s_i=
1$ with probability 1/2 and $s_i$=0 with probability 1/2. The
voters are aware of the signals of other voters in their community
and are influenced by them.  Let $q>0$ be a small real number. A
voter changes his mind if he observes a decisive advantage for the
other candidate in his community, i.e., if he observes an
advantage where the probability of observing such an advantage or
a larger one, when voter behavior is independent and uniform, is
at most $q$. (We can even assume that a voter only sees the outcomes of
an election poll and also that only a small fraction of voters are
influenced by the views of others.)

The election's outcome as a function of the original signals
$s_1,s_2, ... ,s_n$ can be described by a Boolean function which
we denote by $G[a;b;q]$.

Let us examine the situation for a sequence $(f_n)$ of such
examples where the parameters $a$ and $b$ both tend to infinity,
$n=ab$, $q$ tends to zero and $(1/q)$ is  $o(m)$. (For example,
take $a=b=\sqrt n$, and $q= n^{-1/4}$.) In this case, $f_n$
exhibit noise-sensitivity for (independent) small amounts of noise
in the original signals. The outcome of elections as a function
of the individual signals is thus noise-sensitive.

On the other hand, this same sequence is extremely noise stable
for independent noise with respect to counting the votes! The gap
between votes cast for the two candidates behaves like 
$b\sqrt{qa}$,  so that even if a random subset of
40\% of the votes are
miscounted the probability that the election's outcome will be
reversed is extremely small.

The two properties of this example --- noise sensitivity
for noise affecting the original signal and strong
stochastic stability for noise affecting individual votes --- seem
characteristic to situations in which voters' behavior depends on
independent signals in a way that creates positive correlation
between the voters. Note that when we consider random independent
noise in the original signals,
the distribution of resulting votes is identical to the original
distribution without the noise. This is not the case for random
independent noise in counting the votes.

If we do not know the internal mechanism for creating
the distribution of votes
then the noise looks like some mysterious mechanism that ``conspires'' to foil
the outcome.

It appears that the tensor-product
model of noise is analog to noise in counting the votes
where noise stability is more likely,  but perhaps not sufficiently general.
We have to worry about noise that is more related to the mechanism
that creates the probability dependencies in the system.

%\section {General observations and relevant mathematical issues}

%\subsection {Stability and sensitivity}

%\subsection  {An example concerning elections}

%The issue at hand is about noise sensitivity in systems with
%probability dependence. 
%We consider  a lovely example of an
%election scenario,   taken from [Ka] where small random noise
%looks as if conspiring to spoil the result. 
%(Of course, this
%is only an analogy to the case at hand.) 
%For more details see
%Section \ref {s:election}.

%\subsection {Boolean functions}

%We look at Boolean functions and their Fourier expansion as analog
%objects to noise operators and their expansions in terms of Pauli
%operators. We also consider related hypothetical forms of
%noise. Some related questions and issues can be found in Section
%\ref {s:maj}.

%\subsection {Error-correction}

%The question whether error-correction is possible under more
%general type of noise is a very interesting outstanding problem.
%An approach toward some of
%the conjectures above (and the larger issue at hand),
%in the tradition of
%complexity theory  would be to forget about the quantum model, and consider
%the power of a model with probability distributions supplied (repeatedly)
%by an oracle, subject to a certain type of noise.
%However, error-correction should still take into consideration
%subtle properties of the quantum model. For some related remarks
%see For more details see Section \ref {s:correction}.

\section {Restriction of states of $n$ qubits in noisy quantum
computers}

Here we look at quantum computers from a
different angle. Rather than
thinking about the noise, we consider what is the hypothetical
effect of the noise.

Let us assume that the noise itself is infinitesimally local and
also approximately local. Consider the operator  $S$ which
describes the state of the $n$ qubits at some time along the
computation of a noisy quantum computer. (It is better to think
of $S$ as a random variable). Let $W$ denote the class of these
operators. Let us consider again Conjecture 4.1. This conjecture
asserts that there is some sort of a correlation between the
entanglement of the noise and the entanglement of the signal. We
referred to the elections example (Section 5.2) to suggest that
such a relation can be possible and damaging. However, this
example relies on the noise and the signal both have similar
structure and depend on the same ``hidden'' signals.

In our models the noise is infinitesimally local and stochastic.
Let us examine how the noise can ``conspire'' against the computation.
How would the noise ``know'' what would be the entanglement
in the signal involving a large set of qubits? One explanation would be that
the space of operators describing the state of the qubits is very confined
in order that:
\begin {itemize}
\item 
%[(6.1)]
The dependencies between
large sets of qubits are determined by the 
dependencies  between small sets of qubits --- and in a similar way
for the noise and the signal.
\end {itemize}

This line of thought suggests that in order for the noise to be
damaging as we expect it to be, the possible states of the
$n$-qubits of our computer should be very limited. The following
bold conjecture is in this direction.

\begin {conj} For the case of realistic infinitesimally
local and approximately local noise, the class $W$ of operators
$S$ representing states of noisy quantum computers on $n$ qubits is
confined:
\begin {itemize}
\item [(a)] S itself is approximately local.

\item [(b)] (stronger) $S$ can be written as: $S$ is equal to $S_1 +
S_2$ where

(*) $S_1$ is 
up to classical operations, 
an approximately local
and infinitesimally local stochastic operator.

(**) $e(S_2)$ is uniformly bounded.
\end {itemize}
\end {conj}

Conjecture 6.1(a) asserts that for a realistic approximately local noise 
the operators representing the
states of the $n$ qubits in a noisy quantum computer are
approximately local. The stronger part (b) asserts that
essentially all that  can be done in quantum computers apart from
operators in $L(k)$ for bounded $k$ is to apply classical gates to
an initial state described by a stochastic infinitesimally local
operator. (Since there is no canonical 
way to embed classical computation in the quantum model, the 
term ``up to classical computation'' is concrete only in terms of complexity.)
%It is possible that a stronger conjecture is true, namely that 
%$S_1$ is actually an infinitesimally local stochastic operator'')

An even stronger version would say that $S_1$ is just a 
noise operator and that even classical operations on such operators 
cannot be maintained.

\begin {conj} The class $W$ of operators
$S$ representing states of noisy quantum computers on $n$ qubits is
confined:
$S$ can be written as: $S$ is equal to $S_1 +
S_2$ where $e(S_2)$ is uniformly bounded and
$S_1$ is noise.
\end {conj}

Some stronger versions may suffice to reduce noisy
quantum computers to classical ones.
Conjectures 6.1 and 6.2  represent the most optimistic form of the
pessimistic direction
concerning quantum computers: namely, we can take the complexity away and
make a time-free statement on the limitation of quantum computers.
If true, such a statement under suitable assumptions concerning the noise 
may yield to a proof that is inductive
on the quantum circuit. This direction is worth trying.
(Replacing the absolute bound on $k$ by a slowly growing function of $n$
like $\log n$ may still be useful.)

In this context the work of Aharonov, Ben-Or, Impagliazo, and Nisan [ABIN]
is relevant. They considered the model of noisy reversible
computation and showed that it can be reduced to quantum
computation of depth $O(\log n)$. The strength of the general model
compared to the reversible model lies in the ability to regain 
entanglement between qubits by
extending a certain ``restriction'' of the $n$ qubits to a subset of
the qubits  using fresh qubits. 
We want to argue that a correlation
between the entanglements of qubits in the computer and the
entanglements between the noise operators acting on them will
maintain the restriction of the states of the computer, and, in
particular, will force the state of the computer to be
approximately local if the noise is.

{\bf Remarks:}
%1. Conjecture 6.1 comes with an obvious  
%potential counter example:
%$$ 1/\sqrt 2 <000\dots 0| + <111\dots1|.$$ 
%If this famous quantum position can be practically 
%realized for large $n$ (and there are some claims it was already 
%been realized for as many as 64 qubits which is large enough)
%this will be a reason to send Conjecture 6.1 back to the drawing 
%board or worst. 

1. Greg Kuperberg pointed out that the idea (which he 
regards as baseless) that the quantum states for 
large quantum computers (or for complex quantum systems in nature) 
are confined is not new, and is referred 
to as ``censorship'' in the physics literature. (Of course, 
complexity theory gives very severe (but elusive) 
forms of ``censorship'' both in the classical and in the quantum case.) 
A paper by Aaronson
[AA]
studies the power of quantum computation under several forms of 
censorship. Aaronson attributes the forms of ``censorship'' he considers 
to breakdowns of the laws of quantum physics for large systems.
(Such a possibility was considered by several people, see e.g., Levin [Le].)
In my opinion, much more interesting reasons for ``censorship''
would come from  mundane properties
of noise, well within the laws of quantum physics. 
%In any case, Aaronson's 
%analysis of computational implications of various forms of censorship
%does not depend on their origin.

%\footnote {One 
%speculation for censorship that Greg mentioned is that it results from 
%breakdown of the laws of quantum physics for large systems. However,
%in this paper I am only interested in reasoning 
%which fits the laws of quantum physics.} 

2. I do not know
if there are situations in nature in which entanglement cannot be
regarded as approximately local. (Indeed, successful quantum
computing appears to rely on such scenarios.) It might be possible for
infinitesimally local noise to lead to noise operators that are not
approximately local, and in such
cases we could expect the operator describing
the state of the $n$ qubit not
to be approximately local either. I would expect that in the presence
of such a noise no form of computation is possible and that the
not approximately local component of the operator describing the state
of the quantum computer in such a case is just noise.

%2. It is not clear at all that $L(1)$ -   tensor product operators
%is the correct place to start the filtration of the space of
%operators as far as decoherence is concerned. Decoherence can
%effect systematically "classical" probability distributions and
%destroy also classical forms of probability dependence. Thus the
%correct "fixed point space" may be a much smaller set than L(1),
%and so an even more drastic view would be that even any
%non-trivial classic probability dependence among (many) qubits of a
%quantum computer represents noise.
%For example, we can replace $L(1)$ by $M(1)$ the space product
%distributions on the individual qubits and let $M(k)$ can be the
%product distribution of arbitrary distributions on blocks of at
%most $k$ qubits. 

3. It is known that when we consider ordinary
randomized algorithms it confers no advantage to aggregate with the
random bits throughout the algorithm rather than sample them right
away. Conjecture 6.2 suggests 
that sampling the random bits up-front is essentially the only
method that will work in noisy computers and that even classical 
correlations cannot be maintained along noisy computation. (The only way 
I can think of formalizing such a claim is by considering fragments of 
the quantum model that capture the power of probabilistic 
classical computers.) 

%4. A nonlinear measure of entanglement and an exotic forms of
%decoherence can be found in Section \ref {s:decoherence} .
Our censorship proposals are based on Section 4. We can ask 
what kind of censorship can be expected by the direction of Section 5.1.
Refer by the majority operator to the linear extension 
of the majority function on 0-1 states.

\begin {prob} 

What could be the possible states of a quantum computer 
equipped with a noiseless  majority operator 
subject to noise
considered in Section 3 
\end {prob} 

%\section {Further discussion}

%1. 
\section {Summary of Problems}

A sequence of $\epsilon$-noise operators $T_n$ is {\it devastating}
if in the expansion of $T_n$ to multi-Pauli operators
the weight $w(T_n)$ of multi-Pauli operators of 
height $\ge 0.74 n$ is at least $\epsilon /2$. The sequence
is {\it alarming} if we witness a power-law decay in height of 
weights of multi-Pauli operators. 

\bigskip

\noindent
{\bf Malice:}
\begin {itemize} 
\item[1.] 
Show that for every $\epsilon >0$ 
a polylogarithmic-depth polynomial-size malicious quantum computer can create
a devastating $\epsilon$-noise.
\item[2.]
What is the smallest growth rate of $k=k(n)$ so that 
for every constant $\epsilon >0$, 
there is a devastating noise operator $T_n \in L(k(n))[\epsilon]$  
acting on $n$ qubits? 
\end {itemize}
{\bf Stochastics:}
\begin {itemize} 
\item[3.] 
Show that for every $\epsilon >0$, a random $\epsilon$-noise 
operator $T=T_1 \cdot T_2 \cdot T_m$, 
with $m=\log n$, $T_i \in L(k)[\epsilon']$ ($\epsilon'$ chosen accordingly),
when $k$ grows to infinity with $n$ arbitrarily slow,
is devastating. What is the situation when $k=2$?
\item[4.]
Study the Ising-like model of noise on graphs.
Can it lead to a devastating $\epsilon$-noise? 
alarming $\epsilon$-noise? 
\end {itemize}
{\bf Geometry:}
\begin {itemize} 
\item[5.] 
Show that a devastating behavior in items 2-4  will 
continue to hold under reasonable restrictions based on the 
geometry of the computer.
\end {itemize}
{\bf Conspiracy:}
\begin {itemize} 
\item[6.] 
Describe a model (consistent with the laws of physics) in which classical 
computing is noise-free and quantum computing is noisy.
\item[7.]
Propose non-linear inequalities for decoherence that amount to 
decline of correlations. Show how such inequalities can be derived from 
infinitesimally local behavior.
\end {itemize}
{\bf Majority:}
\begin {itemize} 
\item[8.] 
Show that a majority-based correction can lead to fault-tolerant systems immune
against devastating stochastic noise of the kind considered in Section 3, 
and that no analogous methods are possible in the quantum case.
\end {itemize}
{\bf Censorship:}
\begin {itemize} 
\item[9.] 
Is censorship consistent with the laws of physics? Can it be the 
outcome of mundane properties of noise/decoherence? 
%Find 
%non-linear inequalities for decoherence that imply censorship.
%Explore censorship for noisy computers equipped with noiseless 
%majority. 
\end {itemize}

\section { Conclusion}

The working hypothesis of this paper is that the computational
advantage of current fault-tolerant quantum computation accounts
for the ``classical'' restriction of the noise, and will be reduced
or even completely diminish for other models of noise. 

Adopting and exploring such a pessimistic hypothesis is well in the tradition 
of the theory of computation. Theoretical Computer Science
is famous for its ``pessimistic'' point of view, and there are plenty of
attacks on other computational models based
on worse-case scenarios, powerful adversaries, Byzantine generals,
cryptographic attacks, etc. (This paper has some flavor of a
cryptographic attack.) 
%There are also several results of this nature
%concerning certain fragments of quantum computation.
Such attacks are important
for a theoretic understanding of distributed computation, randomness
in computation, cryptography, 
and various other areas. The mathematics involved
and developed in these studies is often quite exciting and usually
easily recycled. 

%%Similarly, a critical look of the notion of
%%quantum computers which are computationally superior,
%%may raise worthy questions and issues 
%%in complexity theory and mathematics.

%I have not been able
%to find evidence for this hypothesis
%I conjecture also that
%noise which enable any form of computation will  confines the
%states of the qubits of a quantum computers to a "neighborhood"
%($L(k)$) of classical physics.
%and as the reader surely have noticed nothing have been proved here.

The first issue to examine, in my opinion, is how damaging 
infinitesimally local stochastic noise operators can be. 
Finding an alternative model 
to the hypothesis of FTQC, 
that is consistent with the laws of physics,
in which quantum computers are noisy and classical computers are noise-free,
is another interesting problem.

Why noise at all? We took it for granted in this paper, and it
appears to be a clear insight of experts that quantum systems are
noisy. 
%(In fact, we implicitly assume here that this is a basic
%physics phenomenon rather than an engineering difficulty.)
Specifically, it appears to be a common view that the amount of
noise in a quantum computer will be a substantial fraction of the
number of qubits. 
%(E.g., [FKLW] refers to a noise rate of $10^{-4}$
%as ``fantastically low'', while in ordinary (digital) computers the
%noise rate is many orders of magnitude smaller.) 
While it appears
to be clear to experts that quantum systems are necessarily noisy
I am not sure there is a good explanation why this is the case.

And is it correct to think of decoherence as noise?
Perhaps decoherence
is a fundamental property of complex quantum systems that will
remain invariant no matter what physical gadgets are used and which
sub-gadgets are declared to be the qubits --- implying that methods
to eliminate decoherence (error-correction, decoherence-free-spaces,
and even the spectacular topological quantum computers) 
are doomed to fail?

%%In conclusion, quantum computers, weather they can be built or not
%%and quantum computation seem an appropriate mental
%%device to think about complex quantum systems, as well as about
%%computational complexity.
%Part of this work was carried out when the author visited the
%Mittag-Lefler Institute in Djursholm, Sweden. I am thankful to the
%institute and to my host Anders Bj\"orner for the inspiring
%environment. 

\section* {\sc APPENDICES}

\section {Questions on sensitivity and stability}
\label {s:sens.app}
The questions presented here are related to possible connections between 
noise-sensitivity and complexity. 

\begin {prob}
Let $f$ be a $1/\log (n)$-stable Boolean function on $n$ variables 
(or (1/(small $polylog(n)$)-stable). Is it the case that most L-2
norm of $f$ (or a substantial part of the L-2 norm of $f$)
is concentrated on a polynomial number of coefficients?
\end {prob}

If a Boolean function $f$ is $(1/t)$-stable 
then most of its L-2 norm is concentrated
on 
Fourier coefficients of ``levels'' $O(t)$. Showing that if $t=1/\log n$
this implies that most of the L-2 norm of $f$ is concentrated on a
polynomial number of Fourier coefficients is unknown. It is
related to conjectures by Mansour [M] 
and by Friedgut and Kalai
[FK] 
and a work by Bourgain and Kalai [BK]. 
(The techniques used in [BK]
%by Bourgain and Kalai 
may be useful to show that if $f$ is, say, 
$1/\sqrt {(\log
n)}$-sensitive then most Fourier L-2 norm of $f$ is on a polynomial
number of coefficients.)

Another related question is the following:

\begin {prob} Let $F$ be a class of uniformly noise-stable Boolean
functions (not necessarily monotone). (Suppose that for each $f \in
F$ the probability that $f=1$ is 1/2.) Is it true that for some $\delta
> 0$ for every $f \in F$ there are $n$ Fourier coefficients of $f$ whose
sum of squares is at least $\delta$?
\end {prob}

The monotone case is the main Theorem in [BKS]. The
more-than-median-runs function 
($f=1$ if the number of ``runs'' in the 
sequence $x_1,\dots,x_n$  is more than the median 
number of runs when the values of the variables are 
given uniformly at random) gave the motivation for this
question since I expect Fourier coefficients for adjacent pairs will do.

\section {Non-linear relations respected by 
(linear) noise operators: An example}
\label {s:exa}

Consider the following scenarios. We have 3 universal gates $A$, $B$ and $C$
for quantum computing. Suppose that $A$ and $B$ 
enable classic computing but nothing
beyond. Let $W$ be a rather dense set of states for an $n$ qubits quantum 
computer.

For each $w$ in $W$ write an algorithm (applying the gates one by one)
that uses as few $C$-gates as possible and choose the algorithm to be
minimal according to some natural ordering.

The noise is simple: all $C$-gates are defunct; they do nothing.

In this case $N(w)$ is a (deterministic) 
function of $w$ and it is a non-linear function. 

Now, consider a similar scenario where 
the $C$-gates operate with probability 0.8

In this case, $N(w)$ is a stochastic function of $w$ and it is not a linear
function, namely, it is not described by a probability distribution on linear
functions. 

Suppose we use an arbitrary algorithm.
%in the algorithm you still avoid gate $C$ as much as possible but
%otherwise run any algorithm you want.
%d) The same but with b)
%Let w be a state in W and N(w) be what we compute in the noisy process.
In this case
$N(w)$ is not a function of $w$ (alone) but of the algorithm leading
to $w$ that carries more information. Still going from $w$ to $N(w)$
have
a systematic effect which is intrinsically non linear and can be described
by a nonlinear inequality.
In this case we have the non-linear relation (inequality): 
$N(w) \ne w$ if $w$ requires $C$.

{\bf Remark:} Current FTQC do apply when gates are 
faulty with small positive probabilities. The example of this section 
only demonstrates that non linear relations for noise is a possibility.
Showing that there are non linear inequalities that systematically apply 
%for realistic classes of infinitesimally local noise operators 
is a distant goal. 

%To conclude this section we give our wish list for 
%non linear inequalities for decoherence:
%\begin {itemize}
%\item
%States are noisy unless they are classical
%\item
%Dependencies in the noise are positively related to those of the signal
%\item
%Correlations are decaying.
%\end {itemize}

\section {Speculating on non-linear inequalities for decoherence} 
\label {s:decoherence}

%We give here a description of a brute-force measure $D(T)$ of the
%total amount of entanglement (non-linear, complicated and not
%related to error-correction) and a question concerning
%entanglement-reducing noise.

%It seems that we can measure entanglement of a unitary operator
%$T$ on $n$ qubits as follows: Let $T_k$ be the "projection" of $T$
%to the space $L(k)$ consist of tensor products of operators each
%acting on at most $k$ qubits. 

%As $L(k)$ is not a linear subspace the meaning of "projection"
%should be explained. We will choose $T_k$ as the operator in
%$L(k)$ such that $Z_k$, the Hilbert Schmidt norm of $(T-T_k)$ is
%minimal. 

%A measure of entanglement is:

%$D(T) = \sum Z_k$

%$D(T)$ is (a non-linear) analogous to the notion of influence of
%Boolean functions,

%Remark: a similar issue arises when we try to measure the ordinary
%dependence for a certain distributions on random (say Boolean)
%variables $x_1,...,x_n$. Fourier expansions and influences are in
%analogy with the expansion in terms of Pauli operators and the
%quantity $e(S)$. It can be interesting to study a more precise
%(albeit, more complicated) notions.

%Let $S$ be a unitary operator describing the state of the $n$
%qubits and let $T$ be the operator describing the noise.

%\begin {prob}

%1. How large can $D(S)$ be as a function of $n$?

%2. Show that for realistic forms of noise for some $\beta > 0$
%$D(T) > D(S)^\beta$

%\end {prob}

It is interesting to consider 
entanglement-reducing noise
operators, that do not alter states which are
in $L(1)$ to start with.
Noise operators that decrease entanglements are natural from the
mathematical point of view and also from the point of view of physics.
(Mathematically, such operators are related to those studied in
hypercontractive estimates.) From the physics point of view
they are referred to as thermal noise, or thermalization of state, etc.
%\footnote { An ``attack'' on fault-tolerant 
%quantum computing based on thermodynamics considerations can be found in 
%[ALZ].}

A standard simple example is: with some probability
you forget the present state, replacing it
with a unit vector chosen at random (uniformly on the unit sphere, or
equivalently, uniformly from an orthonormal basis).

Basic linear operations of this kind do have
a tensor product form and therefore will yield to current FTQC 
%the Aharonov Ben-Or fault-tolerant 
schemes. (In fact, this is a nice
application of FTQC.)
% their method.) 

The discussion in Section 4 suggests
looking at non-linear relations (inequalities) for 
noise operators that express decreased dependencies between qubits.
Such non-linear relations can be of the following form: we start with a 
class $W$ of correlation-decreasing non-linear operators. The non-linear 
inequality for a noise operator $N$ is that for any state $x$ of the 
computer $N(x)$ is in the convex hull if $T(x)$ for $T \in W$.

A quite natural class of correlation-decreasing 
operators can be obtained as follows.
(This follows a discussion with Yuval Peres and Oded Schramm.)  
Suppose that the coefficients of your distribution are nowhere zero.
Apply your favorite linear thermal noise (like the one from the 
previous paragraph) on the logarithm of the 
distribution, then exponentiate and normalize. (Another way to put it 
is to write the distribution in Gibbs form and apply a linear ``thermal'' 
operator on the exponent.) 

More formally (following Yuval Peres), consider the 
qubits as admitting the values +1
and -1. For a nowhere-zero distribution given by a complex vector of 
length $2^n$
$\mu(x_1,x_2,...x_n)$, $x_i$ = +1 or -1, $i=1,2,\dots,n$,
write

$$\mu(x_1,...,x_n)=e^{-H(x)}/Z,$$ where
$H(x)=\sum_{k \le n} H_k(x)$, and $H_k$ is a
homogeneous polynomial of degree $k$ in $x_1,\dots,x_n$.
Consider the following family D of operators: 
map the measure $\mu$ above to
measures $\mu_t$ that have a similar form, $ \mu_t(x)=e^{-H(x,t)}/Z(t)$ where
$H(x,t)=\sum_k c(k,t) H_k(x)$ with $c(1,t)=1$ and $c(k,t)$ decreasing in $k$ 
and
in $t$, with $c(k,t)$ tending to zero as $t$ goes to infinity for each $k>1$.

As we said, we cannot expect that the decoherence operator will be of such a 
form but rather that for a quantum computer at a state $s$ the 
value $N(s)$ of the (linear) decoherence operator will be in 
the convex cone of dependence-reducing operators like those  
described here.

Such a property of decoherence %if indeed realistic
%(e.g., representing the kind of phenomena Alicki, Horodecki,
%Horodecki and Horodecki propose) 
may amount to an ``elasticity''
behavior with respect to entanglement. When you apply a process leading to an
entanglement there will be some
persistence of or recoil towards the existing state with no effect
in the case of no entanglement.

Another class of operators which I find mathematically appealing
can be described as follows.
Let $V$ be a normed vector space and 
$U_0\subset U_1 \subset \dots \subset U(k)$ 
be a filtration of it.
For $v \in V$ let $v_k$ be the projection of $v$ to $U_k$,
namely, let $\|v-u_k\|$ be minimal among $u_k \in U_k$.
%We will describe some such operators.
%Let $S$ be an operator describing the state of the quantum computer.
For $\epsilon >0$ define
$$N_\epsilon(v) = \sum \epsilon ^k (u_{k+1}-u_k).$$
When the filtration is described by flags of vector spaces
we obtain familiar linear ``contractive'' operations. Operators
of this kind related to other 
filtrations (e.g., filtrations of the space of matrices
according to  rank) look interesting. 
Let $N_\epsilon$ be the operator that corresponds to the filtration
$L(1) \subset L(2) \subset \dots$ of quantum operations.  
%(or more generally, quantum operations.)

\begin {conj}
For every $\epsilon >0$, quantum computation
subject to the operator $N_{1-\epsilon}$ is polynomially reducible
to probabilistic classical computation.
\end {conj}

This (or a somewhat weaker statement) may yield to 
the fundamental simple argument by Aharonov, Ben-Or,
Impagliazo, and Nisan.

\end {document}